\documentclass[aps,notitlepage,a4paper,superscriptaddress,11pt]{article}

\usepackage[utf8]{inputenc}
\usepackage[pdftex]{graphicx}
\graphicspath{/ /.images/ /}
\usepackage[dvips]{color}
\usepackage{breqn}
\usepackage[toc]{appendix}
\usepackage[hang,flushmargin]{footmisc} 
\usepackage{gensymb}
\usepackage{titletoc,ntheorem-hyper,authblk,abstract,latexsym,microtype,booktabs,amsmath,cleveref,fancyhdr,wrapfig,array,amsfonts, amssymb,geometry,appendix,cite}
\usepackage{float}
\usepackage{placeins}

\newcommand{\be}{\begin{equation}}
\newcommand{\ee}{\end{equation}}
\newcommand{\bea}{\begin{eqnarray}}
\newcommand{\eea}{\end{eqnarray}}
\newcommand\blfootnote[1]{%
  \begingroup
  \renewcommand\thefootnote{}\footnote{#1}%
  \addtocounter{footnote}{-1}%
  \endgroup
}

\usepackage{caption}
\usepackage{subfigure}
\usepackage{amsmath}
\usepackage{amsfonts}
\usepackage{amssymb}
\usepackage{stackengine}
\usepackage{mathtools}
\numberwithin{equation}{section}
\usepackage{sectsty}
\sectionfont{\centering}

\numberwithin{subcase}{case}
\usepackage{framed}
\allowdisplaybreaks
\usepackage[nottoc]{tocbibind}
\usepackage[free-standing-units]{siunitx}
\usepackage{helvet}
\usepackage{url}
\usepackage{titletoc}
\usepackage{blindtext}
\usepackage[gen]{eurosym}

\title{Ermakov-Pinney equation for  time-varying mass systems}
\author{Bijan Bagchi}
\author{Achal Vinod}
\affil{Physics Department, Shiv Nadar University, Gautam Buddha Nagar, \\
Uttar Pradesh 201314, India}

\begin{document}
\maketitle
\begin{abstract}
We extend Fring-Tenney approach of constructing invariants of constant mass time-dependent system to the case of a time-dependent mass particle. From a coupled set of equations described in terms of guiding parameter functions, we track down a modified Ermakov-Pinney equation involving a time-dependent mass function. As a concrete example we focus on an exponential choice of the mass function. 
\end{abstract}

\blfootnote{E-mails: bbagchi123@gmail.com,  av927@snu.edu.in}

\section{Introduction}

For some time now, there have been several efforts to study time-dependent quantum systems from different perspectives. One such consists in the use of optical traps (see for some representative articles \cite{ash1, neu, ash2, met, ash3}). In the non-relativistic context, the governing equation of concern is the time-dependent Schr\"{o}dinger equation with the explicit presence of an optical potential. It is usually taken to be in a separable form which is made up of a time-dependent modulation term together with some trapping shape described by a spatial profile \cite{oha, ful}. 

Interest in exploring the dynamics of time-dependent quantum systems started with the method
of seeking Hermitian invariants \cite{lew1, lew2, lai} for them. Recently, Fring and Tenney \cite{fri1, fri2} made a systematic study of time-independent approximations for a class of time-dependent optical potentials. Their general strategy was to write down an approximate Lewis-Riesenfeld scheme \cite{lew2} and explore, among other things, time-dependent potentials with a Stark term. While constructing time-independent invariants, even for non-Hermitian systems \cite{fri2, fri3, mam}, one of the central issues was to derive the highly nonlinear Ermakov-Pinney $(EP)$ equation \cite{erm, pin}.  Interestingly, they could obtain a regular solution to it, determine the electric field and finally construct the invariants. Note that they concentrated only on a constant mass system. For other approaches towards calculating dynamical invariants we refer to \cite{ber}.

Historically, derivation of the $EP$ equation (see \cite{mor} in which some symmetry aspects of this equation were explored) was first undertaken in an early work of Ermakov \cite{erm} while studying integrablity of certain classes of ordinary second-order differential equation

\begin{equation}
    \ddot{x} + \omega^2 x = \frac{\lambda^2}{x^3}, \quad \lambda > 0
\end{equation}
We see that the $EP$ equation describes an oscillator in the presence of an inverse-cube force term. 

Much later, Pinney \cite{pin} provided the solution and gave a complete representation of it in the form

\begin{equation}
    x(t) = \left (Ar^2 + 2Brs + Cs^2 \right )^{\frac{1}{2}}
\end{equation}
where $r, s$ are both functions of $t$ and stand as the independent solutions of the standard harmonic oscillator $\ddot{x} + \omega^2 x = 0$ while $A, B, C$ satisfy the constraint $AC - B^2 = \frac{h^2}{W}$, $W$ being the Wronskian. 

    The purpose of this article is to pursue an analog procedure as that of \cite{fri1} but consider a time-dependent mass instead of a constant one. The variation of the mass with time is a well studied topic in problems of unstable particles \cite{bek, lan} in the realm of particle physics. A modification to the form of the Dirac equation in seeking invariants has also been considered for time-dependent masses in Dirac motivated relativistic systems \cite{zha}. Cases of both time-dependent and position-dependent mass have also been explored in the literature (see, for instance, \cite{ped, bag}). 
    
    In the following, we basically
utilize invariants to solve the time-dependent
Schr\"{o}dinger equation when the guiding time-dependent potential is factorizable\footnote{For a Hamiltonian which is time-dependent such a situation does not in general occur} in terms of a Gaussian profile apart from a time-dependent piece. We see that a modified EP-equation emerges as an auxiliary equation whose form differs from what is obtained in the constant mass case. However, its highly nonlinear nature precludes us from finding any exact analytical solution. We have therefore gone for a specific case-study of the mass profile as given by an exponential representation. It reduces the basic equations to a tractable form thus facilitating a simple numerical assessment.\\

This work is organized as follows:\\

In section 2, we highlight the consistency conditions that emerge when the basic equation for an invariant Hermitian operator $\hat{I}$ is subjected to the Hamiltonian obeying the time-dependent Schr\"{o}dinger equation. In section 3 we derive a modified $EP$ equation by installing  time-dependence in the mass. Such an equation constitutes our subsequent point of analysis which is taken up numerically. In Section 4 we present a particular class of solution corresponding to the exponential choice of the mass. Finally, in section 5, we present a summary.

\section{Constraining equations for time-dependent coefficients }

Let us begin with the time-dependent non-relativistic Schr\"{o}dinger equation which in one-dimension reads 

\begin{equation}
i\frac{\partial }{\partial t} \psi (x, t) = \left [ -\frac{1}{2m} \frac{\partial^2}{\partial x^2} + V(x, t) \right ] \psi (x, t)
\end{equation}
where $m = m(t)$. We work with natural units $\hbar = c = 1$.

For a general quantum mechanical Hamiltonian $\hat{H}$, the time evolution operator $U(t,t_0)$ obeys \cite{sak}

\begin{equation}
i\dot{U}(t,t_0) =\hat{H} U(t,t_0), \quad U(t_0, t_0)=I, \quad t>t_0
\end{equation}
If the Hamiltonian does not depend on time then the solution of the above equation is simply

\begin{equation}
U(t,t_0)= e^{-i\hat{H} (t-t_0)}
\end{equation}
where Hamiltonian is considered to be Hermitian.
 
Against the background of the evolution equation obeying $(2.2)$, we note that the time-dependent invariant operator $\hat{I}(t)$ satisfies 

\begin{equation}
\frac{d\hat{I}}{dt} = \dot{\hat{I}}(t) + \frac{1}{i} [\hat{I}, H] = 0
\end{equation}
where $H$ is time-dependent $\hat{H} = \hat{H}(x, t)$. As a concrete model of our analysis we consider the following representative Hamiltonian

\begin{equation}
\hat{H} (x, t) = \frac{p^2}{2 m(t)} + V(x, t) + x E (t)
\end{equation}
where the mass and the potential have been taken to be time-dependent. The latter is assumed to be factorizable\footnote{For a wider class of time-dependent potentials that are factorizable and also exactly solvable see \cite{spe1, spe2}} in that $V(x, t) = \kappa (t) V(x)$, 
where $\kappa$ is a function of $t$. Additionally, a Stark like term is included in $(2.5)$ which points to the presence of an electric field $E(t)$. The angular frequency $\omega$ is taken to be constant.

The choice of the representation for the invariant is flexible. In the following we adopt the form

\begin{equation}
    \hat{I}(t) = \frac{1}{2}\left[\alpha(t) p^{2} + \gamma(t) x+\delta(t)\{x, p\}+\varepsilon(t) x^{2}\right] 
\end{equation}
where $\{x, p\} = xp + px$ and the coefficient functions $\alpha, \gamma, \delta$ and $\varepsilon$ are time-dependent functions. Of course, these are distinct from $m(t)$ or $\kappa (t)$ and have to be solved for through the consistency equations.

We now address the simplest class of time-dependent Hamiltonian as induced by a varying mass harmonic oscillator potential. In such a case $\kappa$ corresponds to $m (t)$ while $V(x) = \frac{1}{2} \omega^{2} x^{2}$. Substituting in $(2.4)$ we obtain through the use of $(2.5)$ the commutator

\begin{multline}
    [\hat{I},\hat{H}] = \frac{1}{2} \bigg( \alpha m (-\hbar^{2} \omega^{2} - 2i\hbar \omega^{2} x p) + \alpha E (-2 i \hbar) p + \frac{\gamma}{2m} (2 i \hbar) p + \frac{\delta}{2m} (4 i \hbar) p^{2} \\
    + \delta m (-2i\hbar) \omega^{2} x^{2}  + \delta E (-2 i \hbar) x + \frac{\varepsilon}{2m} (4i\hbar x p + 2\hbar^{2} ) \bigg)
\end{multline}
When substituted in $(2.4)$ the following set of conditions is easily obtained

\begin{eqnarray}
  &&  \dot{\alpha} = -\frac{2 \delta}{m} \\
  &&  \dot{\delta} =  \alpha m \omega^2 - \frac{\varepsilon}{m} \\
  &&  \dot{\varepsilon} = 2 \delta m \omega^{2} \\
  && \dot{\gamma} = 2 \delta E 
\end{eqnarray}
We also have an additional relation 
\begin{eqnarray}
   && \gamma = 2 m \alpha E
\end{eqnarray}
along with a closed form for $E(t)$ given by

\begin{equation}
    E(t) = \frac{1}{m} e^{\int Q(t) dt}, \quad Q(t) = \frac{3\delta}{\alpha m}
\end{equation}
We should mention that in the constant mass case, the set of equations $(2.8) - (2.12)$ reduce to those of the corresponding ones in \cite{fri1}.

\section{Ermakov-Pinney equation and its modification}

We aim at writing down a modified EP equation in the presence of $m (t)$. Redefining $\alpha = \sigma^{2}$ the remaining parameters read 

\begin{eqnarray}  
    && \delta = - m \dot{\sigma} \sigma \\
    && \gamma = 2\sigma^{2} m E(t) \\
    && \varepsilon = m^{2} \dot{\sigma}^{2} +  m^{2}\sigma\ddot{\sigma} + m\dot{m}\dot{\sigma}\sigma + m^{2}\sigma^{2}\omega^{2}
\end{eqnarray}
Introducing $\tau \equiv \tau (\sigma)$ as an arbitrary function of $\sigma$ which is defined through
\begin{eqnarray}
  &&  m^{2}\sigma\ddot{\sigma} + m\dot{m}\dot{\sigma}\sigma + m^{2}\sigma^{2}\omega^{2} = \frac{\tau m^2}{\sigma^{2}}
\end{eqnarray}
it is clear that $(3.4)$ stands as the signature of a dissipative  EP equation for the time-varying mass case

\begin{equation}
    \ddot{\sigma} + \omega^{2}\sigma + \frac{\dot{m}}{m}\dot{\sigma} = \frac{\tau}{\sigma^{3}}
\end{equation}
We immediately see that when $m$ is stationary and $\tau$ is a constant equal to $\lambda^2$ the standard equation $(1.1)$ is recovered. 
$(3.3)$ now reads 
\begin{equation}
    \varepsilon = m^{2} \dot{\sigma}^{2} + \frac{\tau m^2}{\sigma^{2}}
\end{equation}
Clearly, no unique set of solutions of the parameter functions exists satisfying $\sigma$-dependent equations $(3.1)$, $(3.2)$ and $(3.6)$, for a given functional choice of the mass function $m(t)$ and the electric field $E(t)$.

Identifying the EP equation in the extended form $(3.5)$ constitutes the central result of this note. Although not exactly coincident with $(3.5)$, a dissipative $EP$ equation was also derived by Fring and Tenney \cite{fri2} but in a different context 

\begin{equation}
    \ddot{\chi} + f^2 \chi - \frac{\dot{f}}{f} \dot{\chi}  = \frac{f^2}{\chi^3}, \quad \chi = \chi (t)
\end{equation}
where $f$ is a time-dependent parametrizing function. $f$ seems to play the analog of the mass function but notice that while the coefficient of $\sigma$ in $(3.5)$ is a constant, the same in the corresponding $(3.7)$ is a function of time.

Our next endeavour would be to analyse $(3.6)$ by taking its time-derivative and comparing with $(2.10)$. Towards this end we obtain the pair of equations

\begin{eqnarray}
    && \dot{\varepsilon} = 2m\dot{m}\dot{\sigma}^{2} + 2\dot{\sigma}\ddot{\sigma}m^{2} + \frac{1}{\sigma^{2}}(2m\dot{m}\tau + m^{2}\tau^{\prime}\dot{\sigma}) - \frac{2m^{2}\tau\dot{\sigma}}{\sigma^{3}} \\
    && \dot{\varepsilon} = -2 m^{2} \omega^{2}\sigma\dot{\sigma}
\end{eqnarray}
where $(3.1)$ is used. Equating them we straightforwardly obtain a differential equation for $\tau$

\begin{equation}
    \tau^{\prime} + 2 \frac{\dot{m}}{m\dot{\sigma}}\tau = 0   
\end{equation}
where the prime denotes a derivative with respect to $\sigma$.

\section{Solving the modified $EP$ equation}

Since $ \tau = \tau(\sigma) $ let us assume a monomial form for it 

\begin{equation}
    \tau = \tau_0 \sigma^{4}
\end{equation}
where $\tau_0$ is a constant assumed to be $> 0$. The EP equation, in principle, admits of complex solutions given by the general form

\begin{equation}
    \sigma (t) = \xi (t) e^{i \eta (t)}, \quad \xi, \eta \in \Re
\end{equation}
where $\xi$ and $\eta$ are linked by a set of coupled equations 
\begin{eqnarray}
    && \ddot{\xi} + \frac{\dot{m}}{m}\dot{\xi} + (\Omega^{2} - \dot{\eta}^{2}) = 0 \\
    && 2\dot{\xi}\dot{\eta} + \xi\left( \ddot{\eta} + \frac{\dot{m}}{m} \dot{\eta} \right) = 0
\end{eqnarray}
where $\Omega^{2}=(\omega^{2}-\tau_{0})$. These are very complicated equations to tackle. So we concentrate on real solutions only and in this regard consider a mass function $m(t)$ of the exponential type

\begin{equation}
    m (t) = m_0 e^{-q t}, \quad q > 0
\end{equation}
where its damping character is in keeping with [21, 22] but with respect to the t-variable. For $\sigma$ we find the solution

\begin{equation}
    \sigma(t) = e^{qt/2}
\end{equation}
subject to the the following constraint on $\tau_0$
\begin{equation}
    \tau_0 = \omega^{2} - \frac{q^2}{4}
\end{equation}
$\tau$ is insensitive to the sign of $q$. For completeness we note that $\delta$ turns out to be a constant while the coefficient function $\gamma$ and the electric field $E$ show a decaying behaviour $\backsim e^{-qt/2}$.

We display the behavior of the mass variation in Figure 1 corresponding to different values of $\omega$. We took the input value of $\tau_0 = 0.01$. The results reflect a typical exponential trajectory of $m$. 
\begin{figure}[h!]
    \centering
    \includegraphics{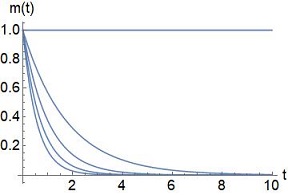}
    \caption{$m(t)$ variation for $\omega=0.1,0.3,0.5,0.7,0.9$}
    \label{f2}
\end{figure}
\vfill

\section{Summary}

In this article we considered the possibility of introducing a time-dependent mass while constructing an invariant for a time-dependent quantum Hamiltonian. Assuming a time-dependent harmonic oscillator with a Stark-like Hamiltonian we were able to derive the invariant conditions in their entirety. We then obtained the time-dependent mass version of the $EP$ equation that included an additional term as compared to the constant-mass case. Because of its strongly nonlinear character we inquired into simple parameter variations to assess the mass function. For the specific case of a damping mass with a certain input value of $\tau_0$, its graphical behavior was plotted over a wide range of $t$.

\section{Acknowledgment}

We are indebted to Andreas Fring and Rebecca Tenney for pointing out errors in the previous version of the draft.

\end{document}